\documentclass[12pt,A4]{article}
\usepackage[english]{babel}
\usepackage{graphicx}

\newcommand{\alt}{\mathbin{\lower 3pt\hbox
   {$\rlap{\raise 5pt\hbox{$\char'074$}}\mathchar"7218$}}}
\newcommand{\agt}{\mathbin{\lower 3pt\hbox
   {$\rlap{\raise 5pt\hbox{$\char'076$}}\mathchar"7218$}}}

\begin{document}

\setcounter{footnote}{0}
\setcounter{equation}{0}
\setcounter{figure}{0}
\setcounter{table}{0}

\title{\large\bf Non-ergodicity effects in 1D localization
 }

\author{\small  I. M. Suslov \\
\small P.L.Kapitza Institute for Physical Problems,
119334 Moscow, Russia \\
\small E-mail: suslov@kapitza.ras.ru\\
{}\\
\parbox{150mm}{\footnotesize \,
It is well-known
that the dimensionless Landauer resistance $\rho$ of an 1D
disordered system obeys the log-normal distribution.  The average
value $\langle \rho\rangle$ for such distribution is not
representative, since it strongly differs from
the typical value $\rho_{typ}$ in a specific sample. In fact,
this conclusion should be revised due to effects of
non-ergodicity. If $L$ is the system size, and $K$ is the number
of realizations of a random potential, then a situation for $L\to
\infty$, $K\to \infty$ depends on the order of  limiting
transitions. If the limit $K\to \infty$ is taken firstly, then
the log-normal distribution is valid for all $L$, if the
condition $\rho\gg 1$ is fulfilled. If the number of realizations
$K$ is restricted, then a situation for $L\to \infty$ is
effectively described by the delta-function distribution, and
$\langle \rho\rangle\approx\rho_{typ}$.  Transformation of the
log-normal distribution can be observed with the use of
experimental technique developed in the context of the universal
conductance fluctuations.  Non-ergodicity effects are essential
for understanding of the difference between the theoretical
predictions for the parameters of the log-normal distribution and
the results of numerical and physical experiments.
  } }

\date{}
\maketitle


\setcounter{footnote}{0}
\setcounter{equation}{0}
\setcounter{figure}{0}
\setcounter{table}{0}

\begin{center}
{\bf 1. Introduction }
\end{center}

It is well-known [1--12], that the dimensionless Landauer
resistance $\rho$ \cite{13,14} of an 1D disordered
system\,\footnote{\,The correct definition of
the conductance of finite systems
is not a trivial issue (see Introduction to papers \cite{10,10a}).
The reason of it is related with a specific feature of the
linear response
formulas: the $\delta$-functions entering them
should be extended to a small width $\gamma$, which is
tended to zero only after transition to the thermodynamic limit;
such procedure is evidently impossible in finite
systems.  To avoid this difficulty, the rather elegant trick was
suggested \cite{13a}:  the finite system is connected to
the massive ideal leads, and the thermodynamic limit is
taken in them.
Realization of this procedure in 1D systems results in two definition
of the resistance (in quantum units $h/e^2$): the Landauer definition
$\rho=|r/t|^2$ (where $t$ and  $r$ are the transmission and reflection
amplitudes), and definition $\rho=1/|t|^2$ by Economou--Soukoulis.
The former corresponds to the four-point measurement scheme, while
the latter to the two-point scheme (see more details in
\cite{10a}).  The second definition is physically less
satisfactory, not providing disappearance of  resistance for the
ideal system; however, it is used by many researches, since it
allows unambiguous generalization to the many-channel case
\cite{10a}.  } is described by the log-normal distribution $$
P(\rho)=\frac{1}{\rho \sqrt{4\pi Dn}}
\exp\left\{-\frac{(\ln\rho-v n)^2}{4Dn}\right\}\,,
\eqno(1)
$$
where $n$ is the number of sites in the 1D chain.
This distribution follows from the evolution equation
with the discrete coordinate $n$ playing a role of time,
which in terms of the variable $x={\rm ln}{\rho}$
and under condition $\rho\gg 1$ has a form
of the usual diffusion equation with the drift velocity $v$
and the diffusion constant $D$. According to Eq.1,
a typical value of $\rho$ increases exponentially
with the system size, which is an observable
consequence of localization of states \cite{0}--\cite{19}.
The distribution (1) (with $v=D$) was obtained firstly
in
the random phase approximation [1--6]
(applicable in the depth of the allowed band),
confirmed numerically near the initial band edge and
in the region of fluctuation states  [7--9], and derived
analytically for arbitrary energies by the present author
[10--12].

The moments of the distribution $P(\rho)$ grow
exponentially,
$$
\left\langle \rho^m \right\rangle \sim e^{\kappa_m n}\,,
\quad \kappa_m=vm+Dm^2\,,
\eqno(2)
$$
with the exponents $\kappa_m$ depending
quadratically on $m$.
One can conclude
that the average value $\langle\rho\rangle$
is not representative, since its behavior differs essentially
from that for the typical value $\rho_{typ}$ in the
specific sample
%
%
%
%
$$
\left\langle \rho \right\rangle \sim e^{(v+D)n}\,,
\qquad \rho_{typ}  \sim e^{vn}\,.
\eqno(3)
$$
In fact, this
conclusion needs
a revision due to non-ergodicity effects.

According to the usual definition, ergodicity means that
the time averaging is equivalent to the ensemble
averaging; in the present case,
instead of time we deal with
the one-dimensional coordinate $n$.
 Let $L$ is the system length
in units of lattice spacing, and averaging occurs over
$n\le L$ with the use of $K$ realizations of a random
potential. It will be clear below, that a situation for
$L\to\infty$, $K\to\infty$ depends on the order of
limiting transitions. If the limit $K\to\infty$ is taken firstly,
then the log-normal distribution is valid for arbitrary $L$, if a
condition $\rho\gg 1$ is fulfilled. If the number of
realizations  $K$ is restricted, then the result for
$L\to\infty$ is independent of $K$ and corresponds to the
delta-function distribution $P(\rho)$,
$$
\left\langle \rho^m \right\rangle \sim e^{mvn}\,,
\eqno(4)
$$
so behavior of the average $\left\langle \rho \right\rangle$
is the same, as for the typical value $\rho_{typ}$.
Transformation of the log-normal distribution to the
delta-function one is in principle observable
with the use of
experimental technique developed in the context of the universal
conductance fluctuations (Sec.6).

\begin{figure}
\centerline{\includegraphics[width=3.5 in]{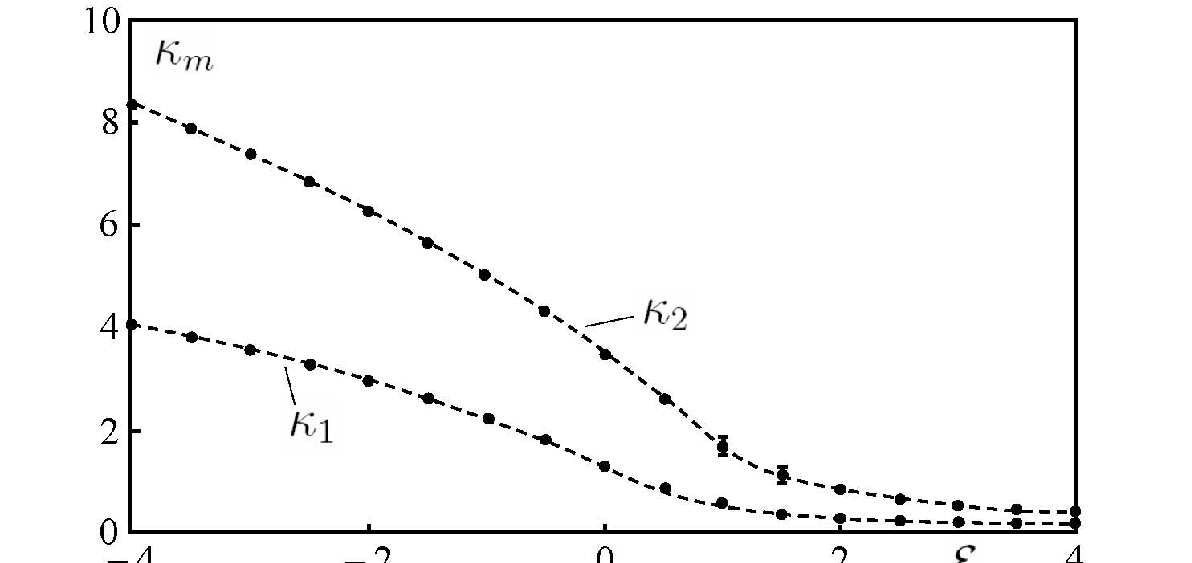}}
\caption{ \small  Theoretical dependencies of the exponents
$\kappa_1$ and $\kappa_2$ (in units $W^{2/3}$) on the energy
${\cal E}$ in units  $W^{4/3}$ (dashed lines) and the results of
numerical experiments (points). To fit by the linear
dependence in formulas (9), the interval $30\le n \le 100$
was used for ${\cal E}\le 1$, while for ${\cal E}>
1$ this interval was extended proportionally to ${\cal E}$;
the latter corresponds to the condition (18), if the dependence
$v\propto {\cal E}^{-1}$ is taken into account.  }
\label{fig1}
\end{figure}

The immediate cause for the present paper was a discrepancy
between theoretical predictions for  parameters
$v$ and $D$ of the log-normal distribution and results of
numerical experiments\,\footnote{\,In the case of appearance of
physical experiments (Sec.6) this discrepancy will lead to
contradiction between theory and the physical experiment.}
(Sec.2). To reveal the origin of
such discrepancy, we undertook  the numerical study of
the moments $\left\langle \rho^m \right\rangle$, corresponding
to the distribution $P(\rho)$. It was established empirically,
that theoretical values are reproduced experimentally for not
very large system sizes $L$ and the sufficiently large number of
realizations $K$. The subsequent theoretical analysis revealed
the striking violation of ergodicity for
power averages $\left\langle \rho^m \right\rangle$
(Sec.3). Meanwhile, the logarithmic averages $\left\langle
\log^m{\rho} \right\rangle$ demonstrate the ergodic behavior
(Sec.4); since the parameters $v$ and $D$ are directly
related with such averages, their behavior is also ergodic.
However, the physical sense of these parameters changes,
when the number of realizations $K$ becomes restricted: for
finite $K$ and $L\to\infty$, the distribution $P(\rho)$
deviates from the log-normal form, and the parameter $D$ does
not characterize adequately its  width. Finally, it was
established  (Sec.5), that the deviations of $v$ and $D$ from
theoretical values is related with different reasons: in the
first case, it is related with renormalization of the energy,
while in the second case it is a direct consequence of
non-ergodicity.

The concept of ergodicity plays a fundumental role in
justification of the statistical physics. The problems related
with existence and violation of ergodicity are extensively
discussed in different fields of physics (see
\cite{100}--\cite{111} and references therein); in the context
of the localization theory they were discussed in
 \cite{107}--\cite{111}.

\begin{figure}
\centerline{\includegraphics[width=3.4 in]{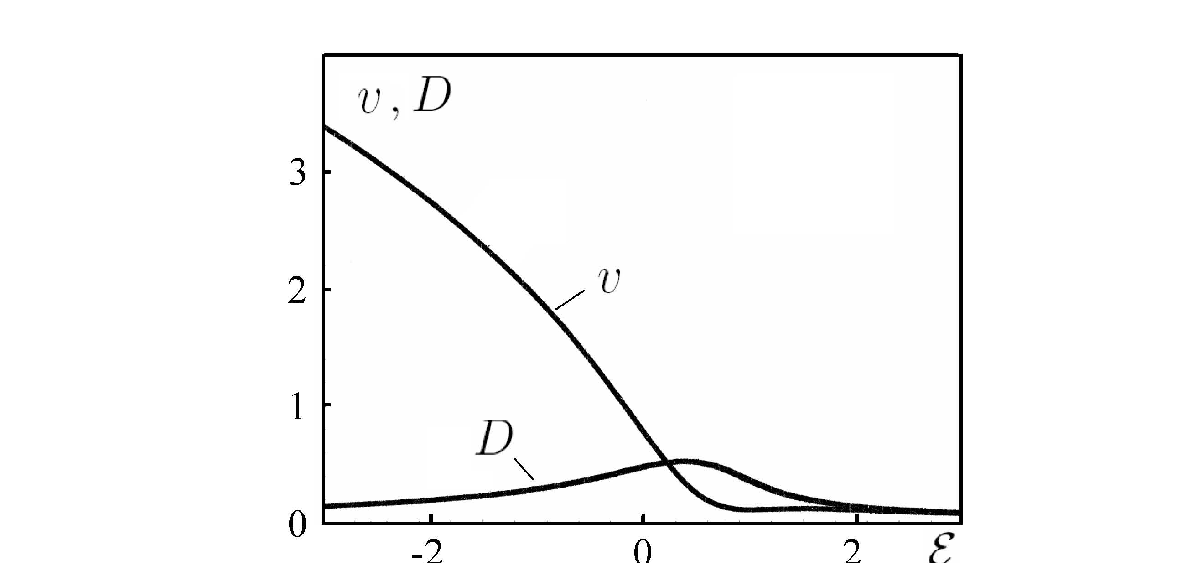}}
\caption{ \small Theoretical dependencies of the
parameters $v$ and $D$ (in units $W^{2/3}$) on the
energy ${\cal E}$ in units $W^{4/3}$.}
\label{fig2}
\end{figure}

\begin{center}
 {\bf 2. Energy dependence of $v$ and $D$ }
 \end{center}

Let have in mind the 1D Anderson model,  describing by the
Schr${\rm\ddot o}$dinger equation
$$
\Psi_{n+1}+\Psi_{n-1}+V_n \Psi_n = E \Psi_n \,,
\eqno(5)
$$
where $V_n$ are statistically independent quantities with
a zero mean and the variance $W^2$, while $E$ is the energy
counted from the band center; all energies are measured in units
of the hopping integral, which was accepted as unity.
Considering Eq.5 as a recurrence relation expressing $\Psi_{n+1}$
in terms of $\Psi_{n}$ and $\Psi_{n-1}$ and setting
the initial conditions on the left end of the system, one can find
the behavior of the second and fourth moments of $\Psi_n$
\cite{10}:  they increase exponentially with $n$, while the
corresponding exponents $\kappa$ are determined by a
positive root of equations
$$
\kappa\left(\kappa^2 +4{\cal E}\right)= 2W^2
$$
$$
\kappa\left(\kappa^2+{4\cal E} \right)
\left(\kappa^2 +16{\cal E}\right)= 42W^2 \kappa^2+96 W^2
{\cal E}  \,
\eqno(6)
$$
with the maximal real part; here ${\cal E}$ is the energy
counted from the lower edge of the band. According to
\cite{10}, equations (6) are valid in the limit
$$
\delta\to 0\,, \quad \epsilon\to 0\,,\quad
\delta/\epsilon^2=const  \,,
$$
where $\delta=k a_0$, $\epsilon^2=W^2/(2k a_0)^2$, $k$ is the
Fermi momentum, and  $a_0$ is a lattice constant; this limit
is realized for weak disorder near the initial band edge and
corresponds to the "white noise" potential
\cite{19}. In terms of ${\cal E}$ and $W$  this limit looks
as
$$
{\cal E}\to 0\,, \quad W\to 0\,,\quad
{\cal E}/W^{4/3}=const  \,,
$$
and allows to use the reduced coordinates, where momenta are
measured in units $W^{2/3}$, and energies in units $W^{4/3}$.
The Landauer resistance $\rho$ is determined by the quadratic
combinations in $\Psi_n$, and the indicated exponents determine
the behavior of $\left\langle \rho
\right\rangle$ and $\left\langle \rho^2 \right\rangle$;
they are denoted as $\kappa_1$ and $\kappa_2$ in correspondence
with Eq.2, and  their behavior is shown in Fig.1.
Since $\kappa_m$ is known for two values of $m$, the use of Eq.2
allows to establish the energy
dependence of  $v$ and $D$  (Fig.2).

\begin{figure}
\centerline{\includegraphics[width=3.2 in]{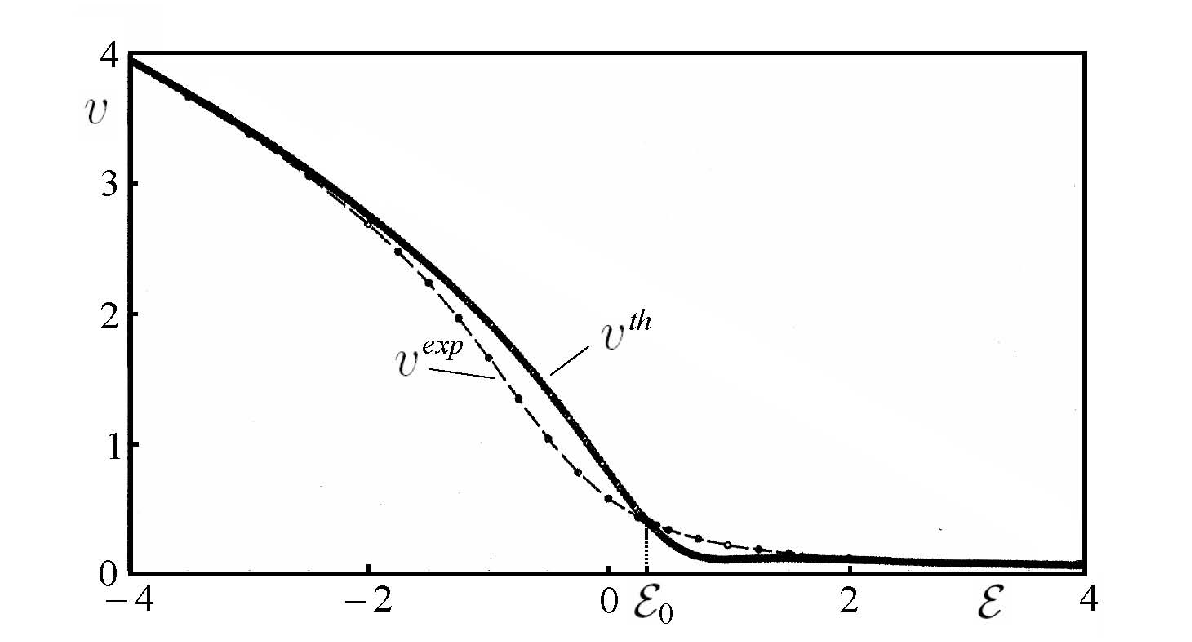}}
\caption{ \small Theoretical and experimental
dependencies for the parameter $v$. }
\label{fig3}
\end{figure}

The cited results can be easily verified in the
framework of the numerical experiment, setting $V_n=Wp_n$,
where $p_n$ is a succession of the Gaussian random numbers with a
zero mean and the unit variance, which can be generated by
the standard procedures \cite{20}; in what follows, we use the
value $W=0.01$.  Such verification led to unexpected results:
experimental values of $v$ and $D$, defining by relations
$$
\left\langle {\rm ln}{\rho} \right\rangle=v^{exp}n+c_1\,,
$$
$$
\left\langle \left({\rm ln}{\rho}-v^{exp}n\right)^2
\right\rangle= 2D^{exp}n+c_2\,,
\eqno(7)
$$
displayed the behavior, different from theoretical (Figs.3, 4).
The Landauer resistance  $\rho$ is determined by the quadratic
form of $\Psi_n$ and $\Psi_{n+1}$, and in a rigorous definition
depends on the external momentum in the attached ideal leads
\cite{10,21}; however, this dependence affects only $c_1$ and
$c_2$. In fact, one can use instead $\rho$ the arbitrary
quadratic form; for definiteness, we used the combination
$$
\rho=\Psi_n^2+\Psi_{n+1}^2\,,
\eqno(8)
$$
which never turns to zero and was
exploiting in many papers.

\begin{figure}
\centerline{\includegraphics[width=3.2 in]{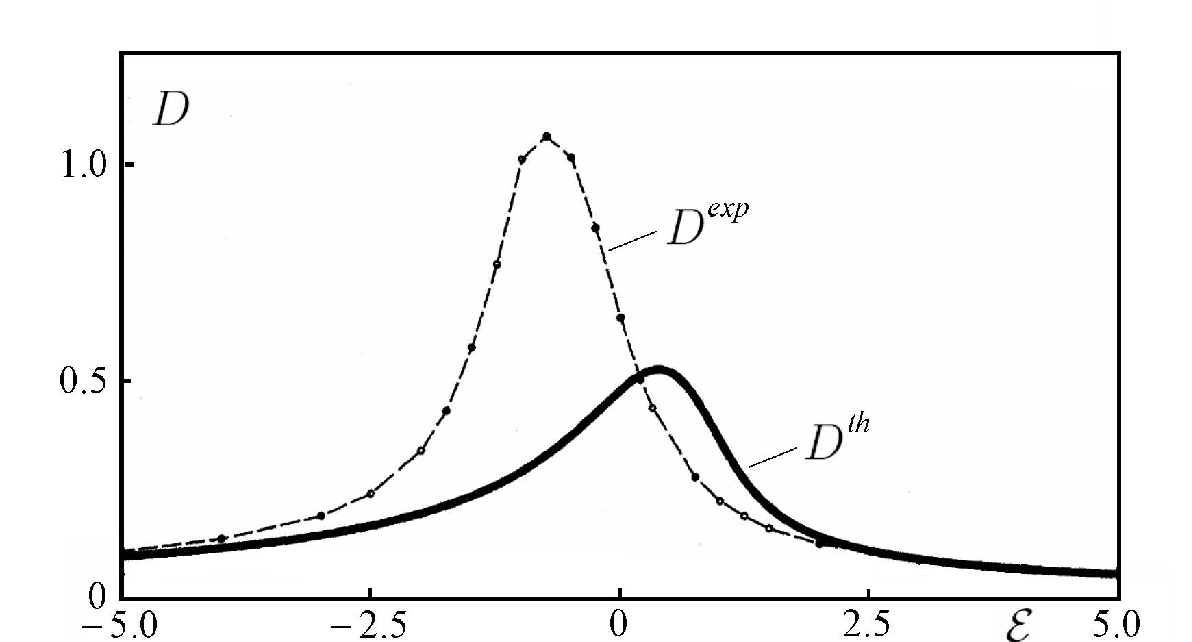}}
\caption{ \small  Theoretical and experimental
dependencies for the parameter $D$. }
\label{fig4}
\end{figure}

\begin{center}
{\bf 3. An origin of deviations from theoretical
values }
\end{center}

To reveal the origin of deviations from theoretical
dependencies we undertook the experimental verification
of the results for $\kappa_1$ and $\kappa_2$, using the
relations
$$
{\rm ln}\left\langle \rho \right\rangle=\kappa_1 n+b_1\,,
\qquad
{\rm ln}\left\langle \rho^2\right\rangle= \kappa_2 n+b_2\,
\eqno(9)
$$
with averaging over $K$ realizations of the random
potential and fitting by the linear $n$ dependence
\cite{20}. If numerical experiments are carrying out
unsystematically, we find it impossible to obtain any stable
results due to their strong dependence on the number of
realizations $K$ and the averaging interval $N_{min}\le n \le
N_{max}$. It was established
empirically, that theoretical results are reproduced
for not very large  $L$ and sufficiently large  $K$.
For the proper chosen conditions of experiment (see the
capture to Fig.1) the correct behavior of $\kappa_1$ and
$\kappa_2$ is reproduced (Fig.1), and consequently
the theoretical values of $v$ and $D$ are
reached for corresponding averaging intervals.

The origin of the strong dependence on  $K$ and $L$ can be
easily clarified. The average $\left\langle \rho^m
\right\rangle$ can be represented in the following form
$$
\left\langle \rho^m \right\rangle=
\int\limits_0^\infty \rho^m P(\rho) d\rho=
\int\limits_{-\infty}^\infty e^{mx} P(x) dx=
$$
$$
=\frac{e^{(vm+Dm^2)n}}{\sqrt{4\pi Dn}}
\int\limits_{-\infty}^\infty dx
\exp\left\{ -\frac{[x-(v+2Dm)n]^2}{4Dn}
\right\} \,,
\eqno(10)
$$
if the change of variables $x={\rm ln}{\rho}$ is performed.
The integrated function $\Phi_m(x)$ corresponds to the
Gaussian distribution, shifted by $2Dmn$ in comparison with
the analogous distribution for $P(\rho)$ (Fig.5). If the shift
$2Dmn$ is sufficiently large, then the actual integration region
corresponds to the tail of the initial distribution $P(\rho)$;
for the restricted number of realizations  $K$,
this region will not contain a sufficient number of
experimental points, and calculation of $\left\langle \rho^m
\right\rangle$ will be incorrect.

If $P(x)$ is the Gaussian distribution with a zero mean and the
variance $\sigma^2$, then the integral over the region $x>s\sigma$
is given by the following estimate for large $s$
$$
P_s=\int\limits_{s\sigma}^\infty\frac{1}{\sqrt{2\pi \sigma^2}}
e^{-x^2/2\sigma^2}\approx \frac{1}{\sqrt{2\pi s^2}} e^{-s^2/2}\,.
\eqno(11)
$$
If the number of realizations $K$ is sufficiently small, $KP_s \alt 1$,
then the region $x>s\sigma$ will not contain even a single
experimental point. If one set
$$
K=e^M \,,
\eqno(12)
$$
then the restriction of $K$ will lead to
an effective cut-off $x_0\approx \sqrt{2M\sigma^2}$
of the Gaussian distribution at the upper limit.
The corresponding cut-off for the distribution (1) occurs
on the scale
$$
x_0=vn+\Lambda\,,\qquad \Lambda\approx \sqrt{4MDn} \,,
\eqno(14)
$$
if the nonzero mean is taken into account. Let introduce
the truncated average $\left\langle \rho^m \right\rangle_{x_0}$,
defined by Eq.10 with the upper limit $x_0$ in the integral.
If $\Lambda\gg 2Dmn$, i.e.  $Dm^2n\ll M$, then truncation is
not essential and one returns to the result (2). In the opposite
case $\Lambda\ll 2Dmn$, the change of variables $x=x_0-y$ gives
$$
\left\langle\rho^m \right\rangle_{x_0}=
\frac{e^{vmn+D(m^2-\tilde m^2)n} }{\tilde m\sqrt{4\pi Dn}}
 \int\limits_{0}^\infty dy
\exp\left\{-y -\frac{y^2}{4D\tilde m^2 n}
\right\} \,,
\eqno(15)
$$
where
$$
\tilde m = m-\frac{\Lambda}{2Dn} \,.
\eqno(16)
$$
The integral is determined by the region $y\alt 1$, where
the $y^2$  term is small due to $Dm^2n\gg M$,
$\tilde m\approx m$, and its neglect leads to the result
$$
\left\langle \rho^m \right\rangle_{x_0} \sim e^{m\tilde v n }\,,
\qquad\tilde v = v+\frac{\Lambda}{n}\approx
v+\sqrt{\frac{4MD}{n}}\,.
\eqno(17)
$$
The diffusion constant  $D$ effectively disappears, while the
positive addition to $v$ arises, which tends to zero for
$n\to\infty$.

\begin{figure}
\centerline{\includegraphics[width=3.2 in]{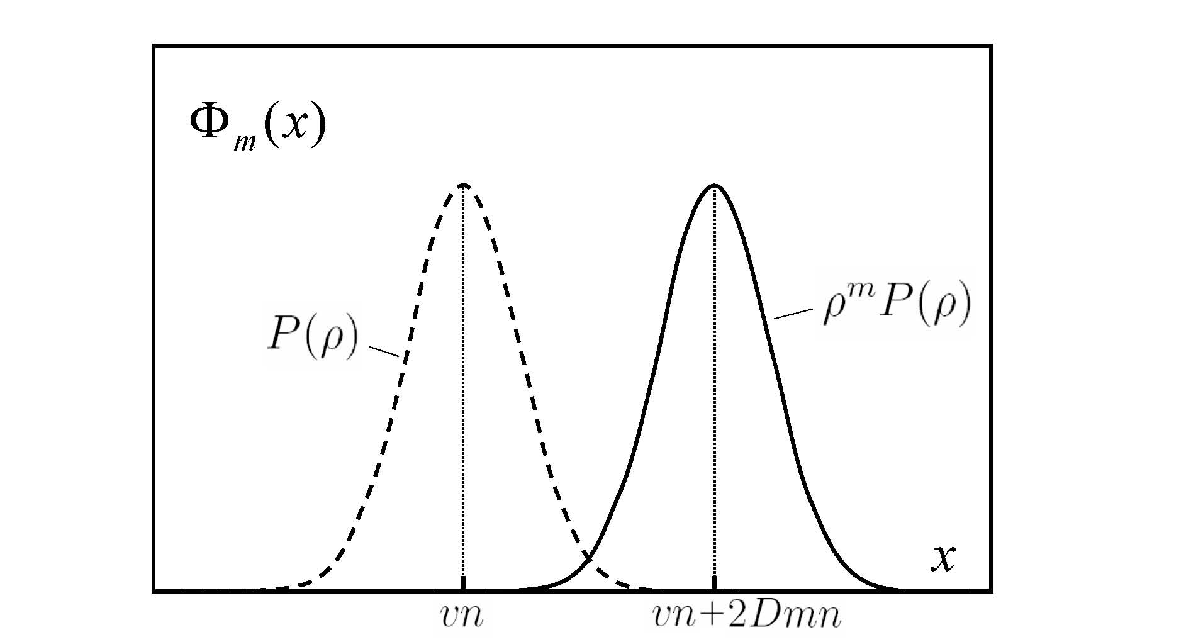}}
\caption{ \small The integrated function $\Phi_m(x)$
in Eq.10, corresponding to $\rho^m P(\rho)$
(solid line) and distribution $P(\rho)$ (dashed line).
 }
 \label{fig5}
 \end{figure}

One can see that correct calculation of the average $\left\langle \rho^m
\right\rangle$ demands the condition $\Lambda\gg 2Dmn$, reducing
to $Dm^2 n\ll M$. Remembering that the log-normal distribution is
valid for $\rho\sim e^{vn} \gg 1$ \cite{4,6,10}, we have
the interval of $n$ values
$$
1\alt vn \alt \frac{M}{m^2} \frac{v}{D}\,,
\eqno(18)
$$
which should be used for fitting to the linear $n$
dependence in the expressions (9);  in fact, the experimental
points in Fig.1 are obtained for the averaging interval
(18), corrected by the visual control for the quality of
fit. The ratio $v/D$ is large in the deep of the forbidden
band (see Fig.2), and the averaging
interval is sufficiently extensive.  On the other hand, $v=D$
in the depth of the allowed band, and $v\sim D$ near its edge;
correspondingly, the allowed interval becomes rather narrow for
$m=2$. For example, we have $M\approx 10$ for $K=10^4$, and
$1\alt vn \alt 2.5$; the increase of $K$ is not very efficient,
since $M$ grows only logarithmically.

To control the validity of the log-normal distribution, one can
use the derivative
$$
\frac{d\kappa_m}{dm} =v +2Dm\,,
\eqno(19)
$$
which is obtained in the result of calculation $\kappa_m$
for two close values $m_1=m+\delta m$ and $m_2=m-\delta m$;
if the same averaging interval is used, and the same set of
realizations is exploited, then diminishing of $\delta m$ is not
related with fluctuations and the loss of accuracy.

The change of the derivative  (19) for the increasing
number of realizations  $K$ is shown in Fig.6:  one can see
that for $K\to\infty$ the linear dependence is extended
to arbitrary $m$, and the full-scale log-normal distribution
arises with theoretical values of $v$ and $D$. The change of
the quantity (19) with increasing of $L$ for the fixed set of
realizations is shown in Fig.7: in the limit $L\to\infty$,
the derivative $d\kappa_m/dm$ tends to the constant
value, and the linear dependence $\kappa_m=vm$ arises,
which corresponds to the distribution $P(\rho)$ close to
the delta-function form.

\begin{figure}
\centerline{\includegraphics[width=3.2 in]{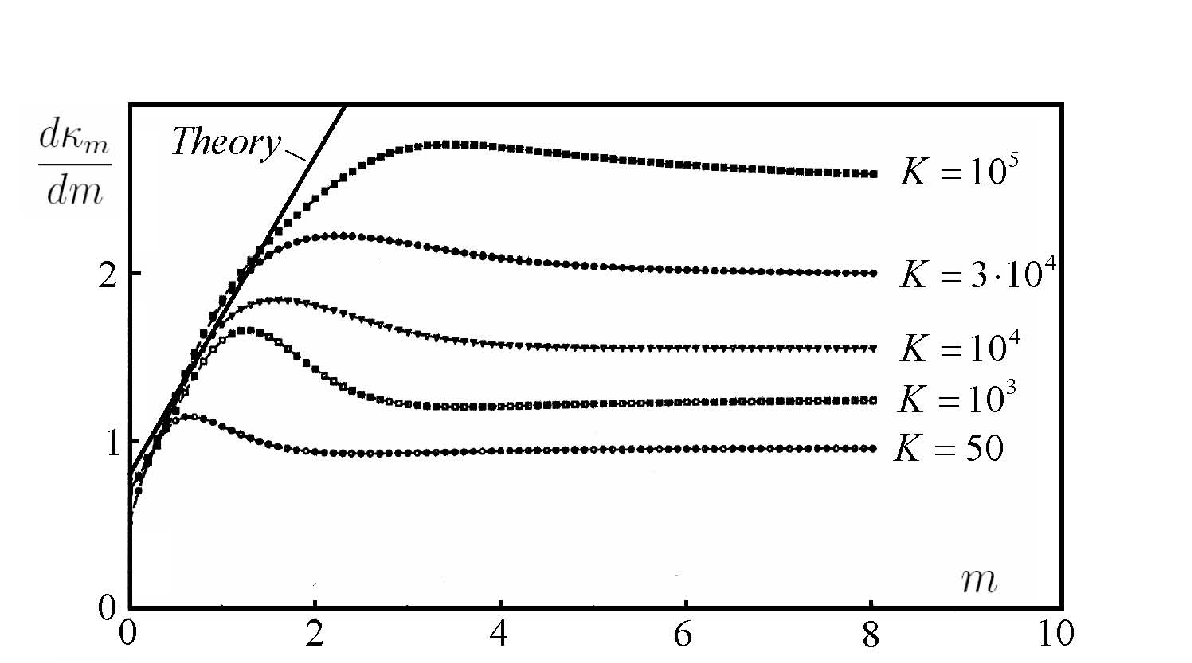}}
\caption{ \small Behavior of the derivative
$d\kappa_m/dm$ at ${\cal E}=0$, $L=100$ for different number
of realizations $K$. }
\label{fig6}
 \end{figure}

\begin{figure}
\centerline{\includegraphics[width=3.2 in]{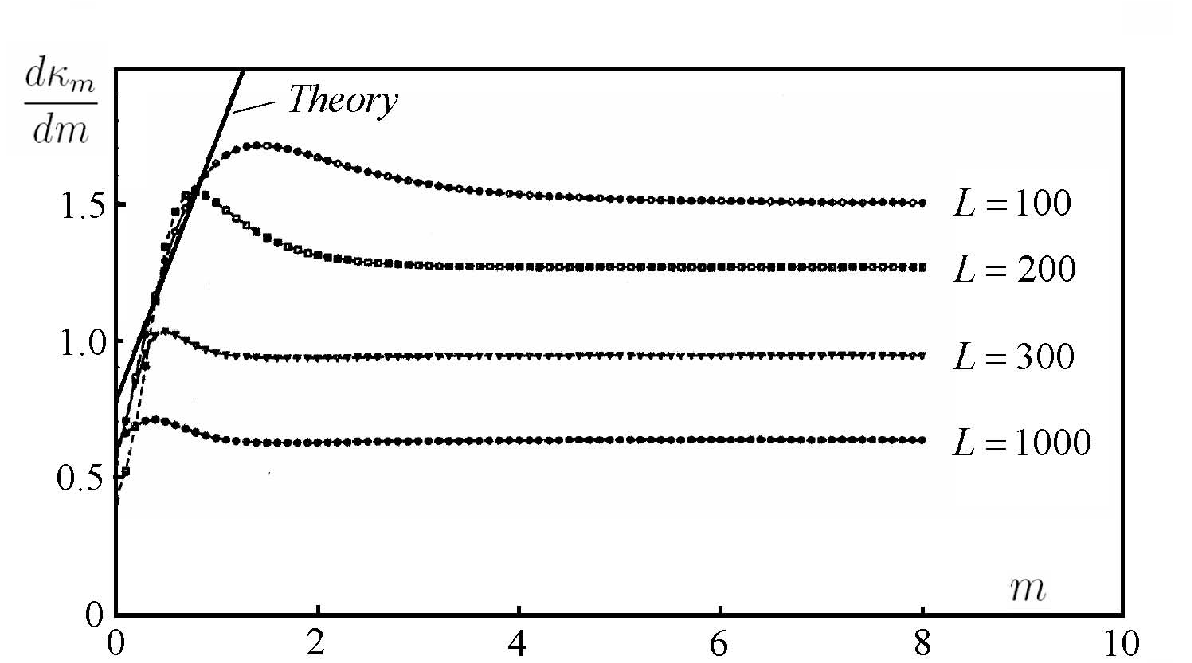}}
\caption{ \small Behavior of the derivative
$d\kappa_m/dm$ at ${\cal E}=0$, $K=100$ for different
system sizes  $L$.
 }
 \label{fig7}
 \end{figure}

\begin{center}
{\bf 4. Transformation of the log-normal
distribution }
\end{center}

The general picture in the $(m,L)$ plane is presented
in Fig.8. The solid line corresponds to the dependence
$L=M/Dm^2$, and in fact is the rough borderline between the
log-normal behavior $\kappa_m=vm+Dm^2$ and the behavior
$\kappa_m=\tilde v m$. If $K$ is increased, then this curve
shifts in the right direction, and for $K\to\infty$
provides the log-normal behavior for all $m$ and $L$.
However, this shift is proportional to the logarithm of $K$
and occurs very slowly. For any practically accessible values of
$K$ a division for two regions is remained.

The boundary between two regions disappears for $m\to 0$;
in particular, it is valid for the logarithmic averages
like (7), determining $v$ and $D$. The estimates of the
parameter $v$ for different $L$ and $K$ (Table 1) demonstrates
complete ergodicity, if  $L$ is sufficiently large ($L\ge 200$).
The limiting value of $v$, achieved for a single realization at
large $L$, is reproduced for all accessible $K$:  the increase of
$K$ leads only to the more rapid achievement of the stationary
value. In fact, Table 1 clearly demonstrates that averaging over
$n$ is equivalent to averaging over $K$: if the interval of $n$
values is increased in 10 times, then one needs 10 times lesser
value of $K$ to obtain the same accuracy.

Estimations of the parameter $D$ (Table 2) demonstrate the
analogous behavior, if the number of realization $K$ is not too
small; in the opposite case one meets with strong fluctuations
with no tendency to a stationary regime. In particular, it
remains unclear, is it possible to obtain the correct value of
$D$ with the use of a single realization.

At this stage one can express a certain scepticism in relation of
the present analysis:  if behavior of $v$ and $D$ is quite
ergodic, then it is sufficient to recover the log-normal
distribution, and calculation of the moments
$\left\langle \rho^m \right\rangle$  is of no
interest.  In fact, it is not so due to following reasons.
In Sec.3 we suggested that
\onecolumn

\begin{center}
{\it Table 1.} \quad Estimates of the parameter
$v$ at ${\cal E}=0$ for different \\system lengths $L$
and a number of realizations  $K$.
\end{center}

\begin{center}

\begin{tabular}{||c|c||}
\hline
$K$      &  $\!\!\!\! L=\, 100 \,\quad
 200 \,  \qquad
10^3 \,\qquad 10^4 \, \qquad 10^5 \,\qquad 10^6
\,\qquad 10^7\,\qquad 10^8 $  \qquad \\
\hline
1 & \,\,\qquad --- \quad  \,2.1239
\, 0.6917  \,   0.6749 \, 0.5900  \,   0.5716 \, 0.5791 \, 0.5790
 \,\,\,\, \\
10 &  0.4804 \,  0.9335 \, 0.5924 \, 0.5844  \,  0.5771 \, 0.5768
\,   0.5796 \,\,\,   \,\,\,\,\,\,\,\,\,\,\,\, \quad \\
$10^2$ &   0.6883 \, 0.4208 \, 0.5610 \, 0.5809  \, 0.5783 \,
0.5798 \,\,\,\,\qquad\qquad\,\,\,  \,\,\,\,\,\,\,\,   \\
 $10^3$ & 0.7459 \, 0.5323 \, 0.5653 \, 0.5771 \,  0.5798
 \,\,\,\,\,\qquad\qquad\,\,\,\qquad\qquad\,\,\,\, \,\,  \\
$10^4$ & \,\,\, 0.7216 \, 0.5760 \, 0.5775 \, 0.5791
\,\,\,\,\,\qquad\qquad\,\,\,\qquad\qquad\,\,\,
\,\,\,\,\,\qquad\qquad  \,   \\
$10^5$ & 0.7260  \, 0.5777 \, 0.5793
\,\,\,\,\,\qquad\qquad\,\,\,\qquad\qquad\,\,\,
\,\,\,\,\,\qquad\qquad\,\,\,\,\,\qquad\qquad  \\
$10^6$ & \,\,\,\, 0.7278 \, 0.5750
\,\,\,\,\,\qquad\qquad\,\,\,\qquad\qquad\,\,\,\qquad\,
\,\,\,\,\,\qquad\qquad\,\,\,\,\,\qquad\qquad\,\,\,\, \\
\hline \end{tabular}
\end{center}

\vspace{3mm}

\begin{center}
{\it Table 2.} \quad Estimates of the parameter
$D$ at ${\cal E}=0$ for different\\ system lengths $L$
and a number of realizations  $K$.
\end{center}

\begin{center}
\begin{tabular}{||c|c||}
\hline
$K$      & $\!\!\!\!\!\! L=\, 100 \,\quad
\,\, 200 \,  \qquad
10^3 \,\qquad 10^4  \,\qquad 10^5 \,\qquad 10^6
\,\qquad 10^7\,\,\qquad 10^8 $   \\
\hline
1 & \,\,\,\,\,\,\,\,\,\,0.9059 \, 1.3197 \, 0.9678 \, 0.3175 \,
0.4508 \, 1.2793 \, 0.9245 \, 0.2036 \,\,\,\,\,\, \, \\
10 & 0.6000 \, 1.1912 \, 0.6417 \, 0.7563 \, 1.3435 \,
1.8249 \, 1.0330 \,\,\,  \,\,\,\,\, \,\,\,\,\,\, \quad  \\
$10^2$ & 1.1080 \, 0.8554 \, 1.0597 \, 1.2573 \,
1.2351 \,  1.2860\,\,\,\,\,\qquad\qquad\,\,\,\,\,\, \,\,\,\,\,\,\,   \\
$10^3$ & 0.7515 \, 1.0846 \, 1.3113 \, 1.2225 \, 1.2844  \qquad
 \,\,\,\,\,\qquad\qquad\,\,\,\qquad\,\,\,\, \,\,  \\
$10^4$ & \, 0.7973 \, 1.2404 \, 1.2836 \, 1.2930 \,\,
\,\,\,\,\,\qquad\qquad\,\,\,\,\,\,\,\,\,\,\,\,\qquad\,
\,\,\,\,\,\qquad \qquad\,   \\
$10^5$ & 0.7834 \, 1.2803 \,    1.2741
\,\,\,\,\,\qquad\qquad\,\,\,\qquad\qquad\,\,\,\qquad
\,\,\,\,\,\qquad\qquad\,\,\,\,\, \,  \\
$10^6$ & \,\,\,\, 0.7792 \,    1.2794
\,\,\,\,\,\qquad\qquad\,\,\,\qquad\qquad\,\,\,
\,\,\,\,\,\qquad\qquad\,\,\,\,\,\qquad\qquad\,\,\,\,\qquad\qquad\\
\hline
\end{tabular}
\end{center}

\vspace{3mm}
\noindent
$P(\rho)$ differs
from the log-normal distribution only by  cut-off of
remote tails, and come to inevitable violation of
ergodicity and its  experimental confirmation
(Figs.\,6,\,7).  However, if ergodicity is violated in principle,
one has no logical reasons to expect that distribution  $P(\rho)$
preserves its log-normal form in the right-hand part of
Fig.5, where theoretical predictions are not reliable.
It can undergo more deep transformations than a simple cut-off
of tails.

The change of $m$ with fixed $L$ and $K$ corresponds to the same
distribution function, and only the averaging quantity is
changed. Let introduce the characteristic function of the
distribution $P(\rho)$
$$
F_\rho(t)=\int d\rho e^{it\rho} P(\rho)\,,
\eqno(20)
$$
which is the generating function of moments
$$
F_\rho(t)=\left\langle e^{it\rho} \right\rangle
=\sum_{m=0}^{\infty} \frac{(it)^m}{m!}\left\langle \rho^m
\right\rangle \,.
\eqno(21)
$$
If all moments $\left\langle \rho^m \right\rangle$ are
known, then Eq.21 defines the characteristic function, while
the distribution  $P(\rho)$ can be obtained by the inverse Fourier
transform. Figs.6 and 7 demonstrate that restriction of the
number of realizations leads to a radical change in the behavior
of moments, which should result in
the essential transformation of
$P(\rho)$; in such case,
the meaning of parameters $v$ and $D$ is also changed.


In fact, the form of the distribution function for large $n$
can be recovered from the experimental data. One can see from
Fig.7, that the derivative $d\kappa_m/dm$ accepts the constant
value $\tilde v$ for $m\agt m_0$, where $m_0$ is
sufficiently small for large $L$. Then one can set
for the integer values of $m$
$$
\kappa_m=\tilde v m +\kappa_0\,,\qquad m=1,2,3,\ldots
\eqno(22)
$$
and receive all integer moments
$$
\left\langle \rho^m
\right\rangle=e^{\kappa_m n}= e^{\tilde v mn+\kappa_0 n}\,.
\eqno(23)
$$

\twocolumn

\noindent
Then substitution to (21) gives
$$
F_\rho(t)=1+\sum_{m=1}^{\infty} \frac{(it)^m}{m!}
e^{\kappa_0 n}
\left(e^{\tilde v n} \right)^m=
$$
$$
=1-e^{\kappa_0 n}
+e^{\kappa_0 n}\exp\left(it e^{\tilde v n}\right)
 \,.
\eqno(24)
$$
The $n$ dependence of $\kappa_0$ is indeterminate a priori,
but the results indicate that the quantity $e^{\kappa_0 n}\equiv a$
is of the order of unity. Then distribution $P(\rho)$ has a
form
$$
P(\rho)=(1-a) \delta(\rho)+
a \delta\left(\rho-e^{\tilde v n}\right)\,.
\eqno(25)
$$
The use of the delta functions is somewhat
conditional, and in fact they should be extended to
a width of the order of unity. In particular, instead of
$\delta(\rho)$ one should use a certain function $P_0(\rho)$,
localized in the region $\rho\alt 1$. Its appearance is
physically explainable. Indeed, in the range $\rho\alt 1$ the
evolution equation for $P(\rho)$ \cite{4,6,10}  is not reduced to
the usual diffusion equation in terms of the variable $x={\rm
ln}{\rho}$; so a small


\begin{figure}
\centerline{\includegraphics[width=3.5 in]{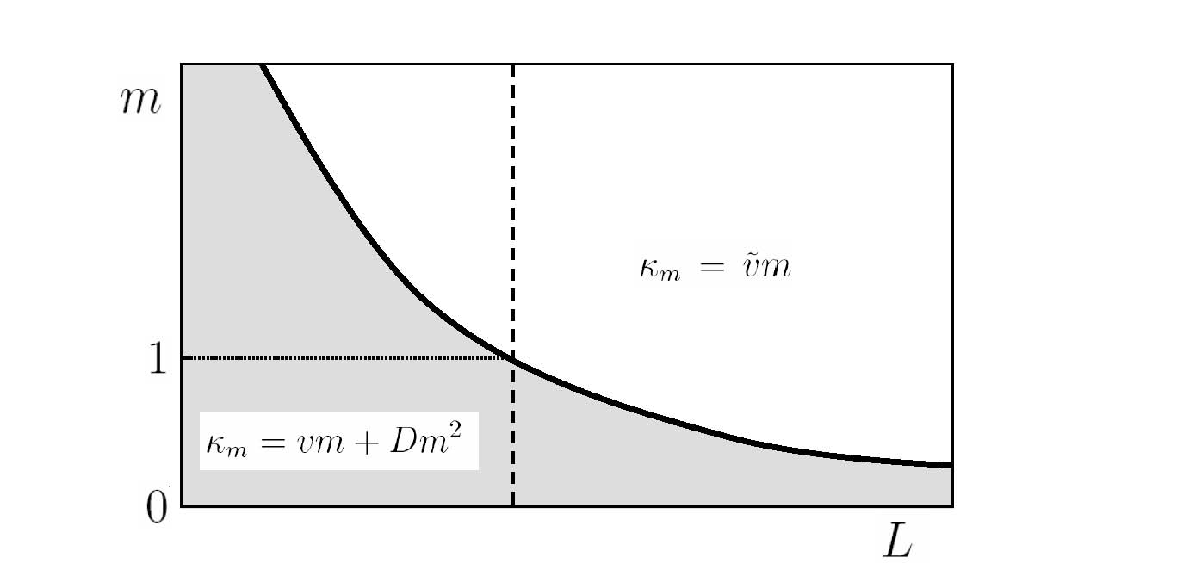}}
\caption{ \small
A solid line in the  $(m,L)$ plane, corresponding to the
dependence $L=M/Dm^2$, is a conditional border-line
between the log-normal behavior $\kappa_m=vm+Dm^2$
and the behavior $\kappa_m=\tilde v m$.
The vertical dashed line is a rough boundary between
the log-normal and delta-function distributions. }
\label{fig8}
\end{figure}

\noindent
part of probability is "stuck"
in the region $\rho\alt 1$ and decreases with $n$
much slowly than in the
interval $1\ll \rho \ll e^{vn}$. After the
change to the variable  $x={\rm ln}{\rho}$ we come to the
distribution
$$
P(x)=(1-a) P_0(x)+
a \delta\left(x-\tilde v n\right)
\eqno(26)
$$
with the characteristic function
$$
F(t)=(1-a) F_0(t)+
a e^{it \tilde v n}\,.
$$
It should be clear from the first line of Eq.10, that calculation
of $\left\langle \rho^m \right\rangle$ is reduced to the change
$it\to m$ in the definition of the characteristic function
$F(t)$, and consequently
$$
\left\langle \rho^m \right\rangle=
(1-a) F_0(-im)+a e^{m \tilde v n} \equiv e^{\kappa_m n}\,.
\eqno(27)
$$
The $m$ dependence of the first term is not essential in
comparison with the strong $m$ dependence of the second term,
and  one can set $F_0(-im)\approx F_0(0)=1$ due to normalization to
unity of the function  $P_0(x)$  in Eq.26. As a consequence, we
have the behavior of $\kappa_m$ for large  $n$
$$
\kappa_m=\frac{1}{n}\, {\rm ln}
\left[(1-a)+a e^{m \tilde v n}\right].
\eqno(28)
$$
Producing expansion for small  $m$, we have
$$
\kappa_m=a\tilde v m +\frac{1}{2} a(1-a)n \tilde v^2 m^2
\equiv vm+Dm^2
\eqno(29)
$$
Since the experimental values of $v$ and $D$ tend to the
constant limits at $n\to\infty$, so it should be
$$
1-a=\frac{2D}{v \tilde v n}\,,
\eqno(30)
$$
and the parameter $a$ tends to unity as $1/n$.
As a consequence, we have the behavior of
$\tilde v$
$$
\tilde v=v+\frac{2D}{v n} \,,
\eqno(31)
$$
which corresponds to the constant value of $\Lambda$
in Eq.17. It signifies that the tails of the log-normal
distribution are cut off not on the scale
$\sqrt{Dn}$, as was suggested in Sec.3, but on the scale of
the order of unity. As a result, the log-normal
distribution transforms to the approximately delta-function
distribution with a width of the order of unity: it corresponds
to effective disappearance of the diffusion constant.
Meanwhile, the experimental value of $D$ does not retain its
initial physical sense and ceases to characterize the width of
the distribution\,\footnote{\,The distribution (26) under
condition (30) has the mean $\langle x\rangle=vn$, while
its variance is determined by the first term and has nothing to
do with the width of the second term. }:  it is simply the
coefficient of $m^2$ in the expansion of $\kappa_m$ for small
$m$.

\begin{figure}
\centerline{\includegraphics[width=3.2 in]{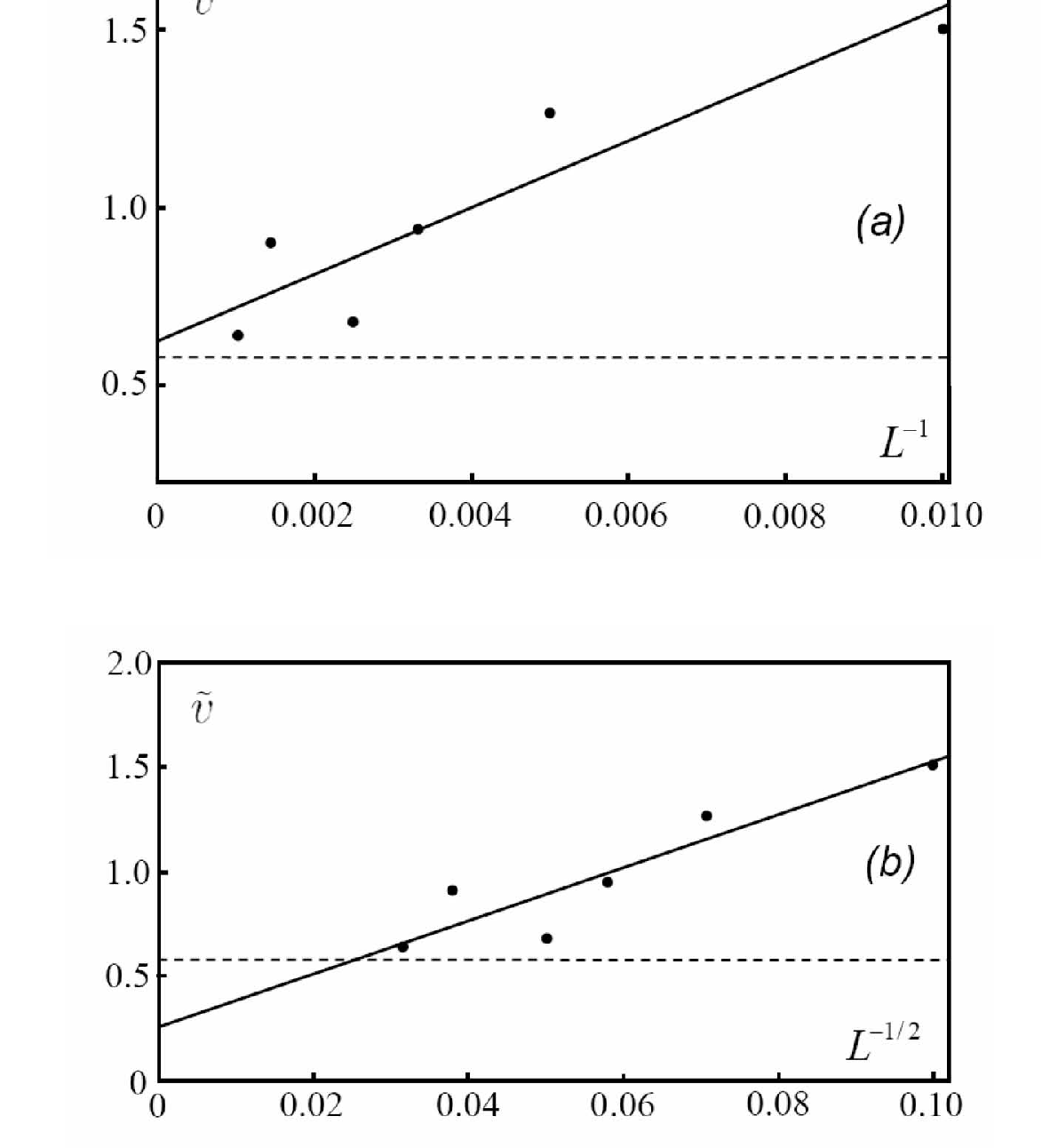}}
\caption{ \small
(a) Extrapolation according to the law
$\tilde v-v\propto L^{-1}$ leads to the value of $v$,
which coincide with that from the Table 1
 (the horizontal dashed line)  within accuracy.
(b) Extrapolation according to the law
$\tilde v-v\propto L^{-1/2}$ leads to the evidently
unacceptable result. The solid lines correspond to fit
according to the least-squares method.  }
\label{fig9}
\end{figure}

\begin{figure}
\centerline{\includegraphics[width=3.2 in]{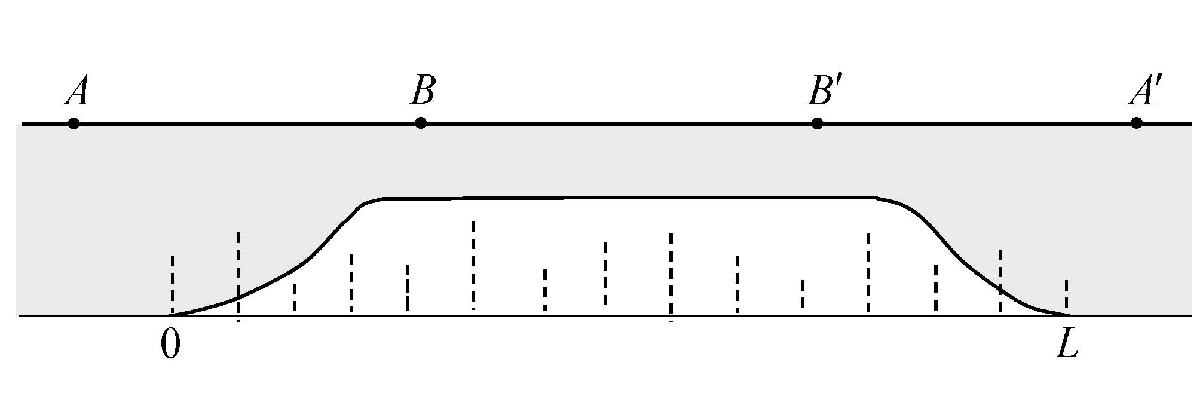}}
\caption{ \small
Due to a fluctuational shift of the allowed band edge in the
deep of the system, the Fermi momentum in the intervals between
the scatterers does not coincide with its value in the ideal
leads. The slow transient behavior arises near the
boundaries of system on the scale of the localization
length $\xi$.   } \label{fig10}
\end{figure}

Extrapolation of  $\tilde v$ to the limit $n\to\infty$ according
to the $1/n$ law is in the reasonable agreement with the value
of $v$ followed from Table 1 (see Fig.9,a). The $L$ dependence
of $\tilde v$ suffers from essential fluctuations and the
quality of fit to dependence $\tilde v-v\propto L^{-1}$
is not very
impressive;
however, attempt of extrapolation
according to the law $\tilde v-v\propto L^{-1/2}$ leads to
the clearly unacceptable result (Fig.9,b).

\begin{center}
{\bf 5. Dependence of $v$ on the length scale and
renormalization of energy }
\end{center}

Experimental estimates of $v$ and $D$ in the small $L$ region
deviate from the general tendency (Tables 1,\,2), and become
close to theoretical values (Fig.6). In fact,
there are physical reasons for  the
dependence of $v$ and $D$ on the length scale.

As was already indicated (see Footnote 1), the correct definition
of the conductance of finite systems demands the attachment of
the massive ideal leads and their explicit inclusion in the
composite system.
Usually, one suggests the "natural" ideal leads made from the
same material, but without impurities. In this case, it is
taken for granted
that the Fermi momentum between the scatterers is the same as
in the ideal leads. This is so indeed for a small number of
scatterers, but a situation is changed when we
come to the
macroscopic system. Existence of the random potential leads to
the fluctuational shift of the allowed band edge, which in the
infinite system corresponds to a change of the energy origin
and has no essential consequences. However, in the
presence of the ideal leads it provides a slow change in the
position of the allowed band (Fig.10), and the Fermi momentum
$k$ is changed slowly  (on the scale of the localization length
$\xi$) from a value in the ideal leads to the constant value in
the depth of the system. For optical systems \cite{12,22}
a situation looks even clearly: if the wave propagates from the
ideal part of the waveguide to its disordered part, then disorder
leads to not only space fluctuations, but  also to a change of the
average refractive index; since the frequency remains
the same, the change of the wave vector occurs.

One can see that a situation is essentially different
for small and large scales. If $L\ll \xi$, then renormalization
of the energy is not essential, and the Fermi momentum coincides
with its value in the ideal leads. Contrary, for $L\gg \xi$ one
can neglect the boundary effects and the value in the depth of
the system becomes actual for the Fermi momentum.  The
theoretical values of $v$ are realized at small scales and hence
are determined by the bare energy ${\cal E}$, while the
experimental values arise at large scales  and are related with
the renormalized energy ${\cal E}_{ren}$; the more formal
arguments for it are given below.

 According to \cite{10}--\cite{12}, parameters of the evolution equations
 are determined  by the Fermi
momentum $k$, being the regular functions of $k^2$, while their
relation with energy is irrelevant. The energy dependencies
of $v^{exp}$ and $v^{th}$ (Fig.3) intersect at the point
${\cal E}_0\approx 0.3$, which was interpreted in Ref.\cite{11}
as a new position of the shifted band edge; so  $k^2$ is equal
to zero. Then the difference between $v^{exp}$ and $v^{th}$
can be explained by the mass renormalization. Let accept
$$
{\cal E}={\cal E}_0+\frac{k^2}{2m_0}\,,\qquad
{\cal E}_{ren}=\frac{k^2}{2m_{\cal E}} \,,
\eqno(32)
$$
where $m_{\cal E}$ depends on energy; then
$$
{\cal E}_{ren}=\left({\cal E}-{\cal E}_0\right)\frac{m_0}{m_{\cal E}}
\,.
\eqno(33)
$$
Theoretical values of $v$ and $D$ are determined by behavior of
the second and fourth moments of $\Psi_n$, which can be
obtained from the Schr${\rm\ddot o}$dinger
equation (5) (see Appendix in
Ref.\cite{10}), containing only the bare energy. Due to
regularity in $k^2$ one has for $v^{th}$ in the small $k$ region
$$
v^{th}=v(k^2)=v_0+v'_0\,k^2=v_0+v'_0\cdot 2m_0 \tilde{\cal E} \,,
\eqno(34)
$$
where $\tilde{\cal E}$ is counted from a shifted
band edge.

The experimental values of $v$ are determined by the log-normal
distribution (1), which can be obtained only with the use of the
transfer matrix in the wave representation \cite{10,11,12},
which does not have a direct relation with the Schr${\rm\ddot o}$dinger
equation\,\footnote{\,In particular, it can be introduced in
optical systems, which are not described by the Schr${\rm\ddot o}$dinger equation.
Such transfer matrix is close to the unit one for a weak scatterer, which
allows to derive the differential evolution equations. The coordinate transfer
matrix, directly related to the Schr${\rm\ddot o}$dinger equation \cite{10},
does not obey such property.}.
If the transfer matrix relates points $A$
and $A'$, situated in the ideal leads (Fig.10), then a slow
transient process will occur in the small $L$ region, related
with the change of the band edge position. To avoid such
transient process, one should use the transfer matrix between the
points $B$ and $B'$, situated in the deep of the system
away from its ends (Fig.10); then the transfer matrix will be
determined by the renormalized energy ${\cal E}_{ren}$. It should
be clear that $v^{exp}$ is the same function of $k^2$

\begin{figure}
\centerline{\includegraphics[width=3.5 in]{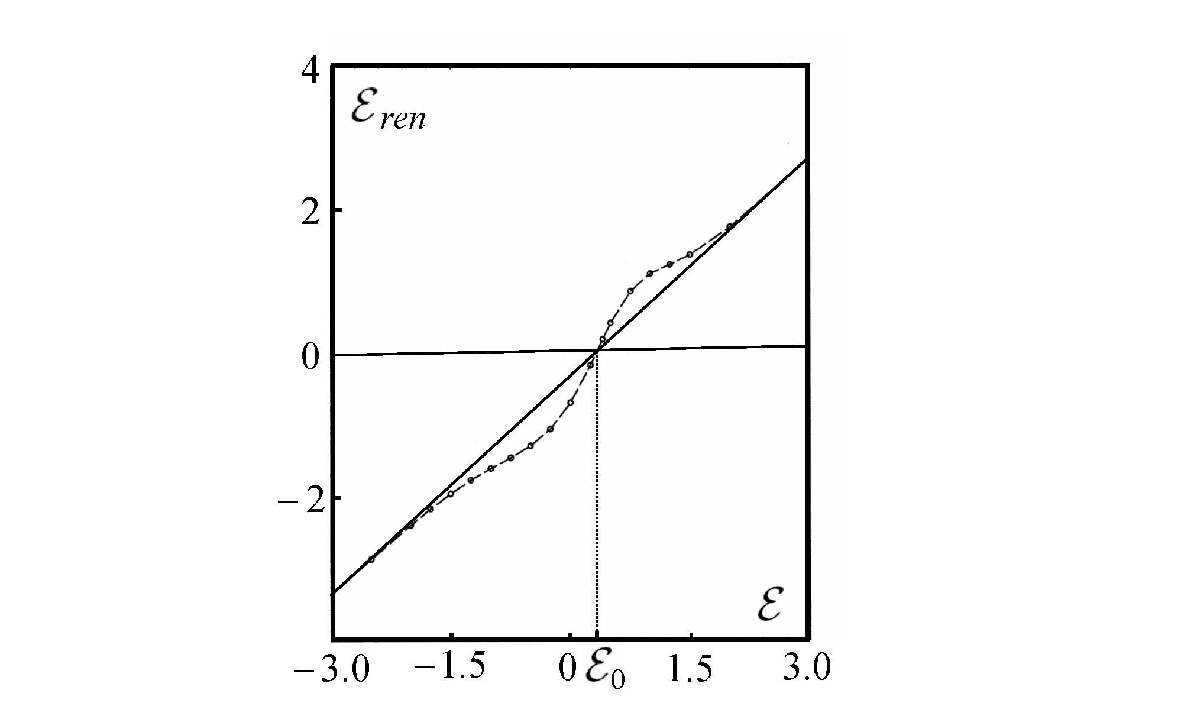}}
\caption{ \small
Relation of the renormalized energy ${\cal E}_{ren}$
with the bare energy ${\cal E}$,following from the difference
between $v^{th}$ and $v^{exp}$ (Fig.3).   }
\label{fig11}
\end{figure}

$$
v^{exp}=v(k^2)=v_0+v'_0\, k^2=v_0+v'_0\cdot 2m_{\cal E}
\tilde{\cal E} \,,
\eqno(35)
$$
but with a different energy dependence. Comparing
(34) and (35), one has
$$
\frac{v^{th}-v_0}{v^{exp}-v_0}=\frac{m_0}{m_{\cal E}}
\eqno(36)
$$
and substitution to Eq.33
gives
the relation between the bare and renormalized energies
$$
{\cal E}_{ren}=\left({\cal E}-{\cal E}_0\right)\,
\frac{v^{th}-v_0} {v^{exp}-v_0}\,.
\eqno(37)
$$
The written expressions are valid for small $k$. However, the
derivative $dv/dk^2$
is almost constant
in the region,
where difference of $v^{th}$ and $v^{exp}$ is essential,
so Eq.37 remains approximately valid. Beyond this region,
the derivative $dv/dk^2$ changes essentially, but then
the mass renormalization is irrelevant
and the corresponding relation $ {\cal
E}_{ren}=\left({\cal E}-{\cal E}_0\right)$
agrees with (37) for $v^{th}=v^{exp}$. Therefore the formula
(37) provides a smooth interpolation between two
regimes, and allows to obtain the relation between ${\cal E}_{ren}$
and ${\cal E}$ (Fig.11),
following from the difference of
$v^{th}$ and $v^{exp}$ (Fig.3).

As was already indicated, the experimental values of $D$
in the large $n$ region do not have physical sense: in fact,
deviation of $D^{exp}$ from $D^{th}$ is induced by
deviation of $v^{exp}$ from $v^{th}$. Indeed, for small $L$
and large $K$ the derivative $d\kappa_m/dm$ has a
well-developed linear portion, which is in agreement with
theory (Fig.6). When $L$ is increased, this linear portion
is shorten,
but the values for $m\sim 1$ remain close
to theoretical.  The experimental values of $v$ and $D$ are
determined by logarithmic averages and correspond to small $m$.
If a value of $v$ for $m=0$ becomes smaller than
the theoretical one, then a slope of curves is increases, and
a value of $D$ becomes
automatically greater; contrary, increasing of $v$ leads
to diminishing of $D$. Such anti-correlation is evident from
comparison of Fig.3 and Fig.4.

The point ${\cal E}_{0}$ was interpreted previously
\cite{11,12,22,24}  as a point of the unusual phase
transition, which is not related with singularities of
resistance $\rho$ and can be observable only in optical systems
\cite{12,22}. If the transfer matrix is introduced between
the points $B$ and $B'$ in the depth of the system, then
with diminishing energy the true
transfer matrix for ${\cal E}>{\cal E}_{0}$ transforms to the
pseudo transfer matrix for ${\cal E}<{\cal E}_{0}$. This
transition can be registered by study  the phase
distribution inside the system, which is allowed by the
modern optical technique \cite{12,22}. Due to spatial
fluctuations of the band edge, the real singularity occurs only
in the thermodynamic limit (as for the usual phase transitions
 \cite{127,128}), when distance between $B$ and $B'$ increases
 unboundedly, and the true position of the shifted band edge
 can be established definitely.

\begin{center}
{\bf 6. Conclusion }
\end{center}

In the usual definition of ergodicity (the time
averaging is equivalent to the ensemble averaging),
it is not concretized, what averaging quantity is kept in mind.
In fact, one can say on the strong ergodicity
(equivalence for all averages), or weak ergodicity
(equivalence for certain averages). As clear from above,
in the case of 1D localization ergodicity is violated for
the power averages $\left\langle \rho^m
\right\rangle$, but remains for the logarithmic averages
$\left\langle \log^m{\rho} \right\rangle$.

To avoid confusion, let us note that in the current literature
strong and weak violations of ergodicity are discussed, which
are understand in the different sense. A strong violation refers
to a situation, when a certain part of the configuration space is
in unaccessible due to infinite barriers (as in spin
glasses). In the case of weak violation, all configuration space
is in principle accessible but there are no equivalence of the
time and ensemble averaging. In the mathematical literature
one introduces ergodicity of the first order
(equivalence of two averages for the first moment), second
order (equivalence for the first two moments), etc.
A certain ordering in the terminology
looks desirable.

The study of effects of non-ergodicity appears essential,
in order to understand the reasons for deviation of theoretical
predictions for the parameters $v$ and $D$ with the results of
numerical experiments. It was established that this deviations
are related with different reasons: in the first case, it is
related with renormalization of energy, while in the second case
it is a direct consequence of non-ergodicity.

\begin{figure}
\centerline{\includegraphics[width=2.8 in]{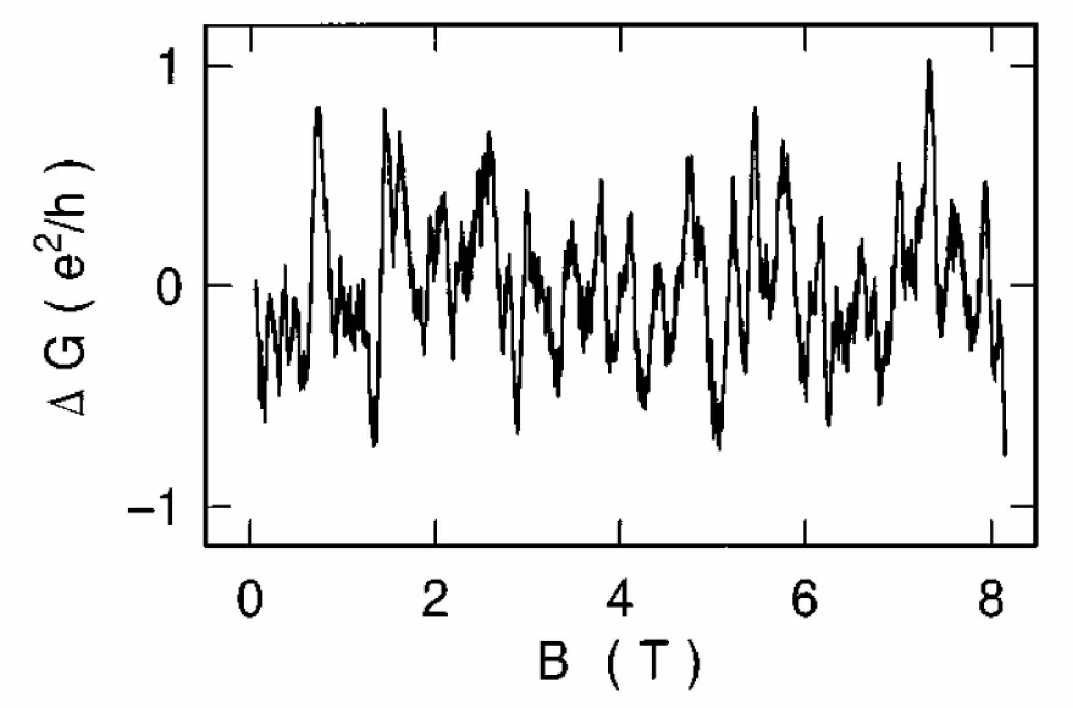}}
\caption{The conductance of the thin Au wire against the magnetic
field  \cite{36}.
} \label{fig12}
\end{figure}

Transformation of the log-normal distribution to
the delta-function one is in principle observable:
the corresponding
experimental technique was developed in the context of the
universal conductance fluctuations \cite{25}--\cite{37}.
The latter are usually observed in the form of aperiodic
 oscillations in the magnetoresistance of thin wires as a
 function of the magnetic field $B$  \cite{36} (Fig.\,12)
(see \cite{600,601} for review).
The fluctuation picture looks random, but is completely
reproducible
in repeated scannings in the magnetic field.
It characterizes a specific realization of the random
potential and changes completely, if the sample is heated till
sufficiently large temperature, at which the impurities become
movable and a new impurity configuration arises.
Performing in such way, one can obtain a sufficiently large set
of different realizations of a random potential in
the same sample. It allows to investigate evolution of the
moments of the Landauer resistance $\rho$, as well as
the total distribution $P(\rho)$.
In particular, the authors of the paper \cite{37}
used  20 realizations of a random potential for study the
evolution of the first two moments of conductance. As clear
from Fig.7, already for $K=100$ realizations effects of
non-ergodicity becomes strikingly manifested.

\end{document}